# THE *G* GRAVITATIONAL PARAMETER AND THE CONCEPTS OF MASS AND DARK MATTER.

Gustavo R. González-Martín

Departamento de Física, Universidad Simón Bolívar,

Apartado 89000, Caracas 1080-A, Venezuela.

Webpage: http://prof.usb.ve/ggonzalm/

The gravitational coupling parameter $G$ is determined by a non riemannian curvature scalar of a background substratum. This substratum represents an inertial solution to the nonlinear equations of a geometric unified theory and provides a limit solution to general solutions in the theory. When a solution approaches the substratum solution as a limit we physically obtain a newtonian limit of the gravitational sector of the unified theory. The curvature scalar is determined by the interaction mass-energy content of the solution. In the limit the equation reduces to Poisson's equation and the curvature scalar reduces to the constant substratum curvature scalar which determines Newton's gravitational constant $G$. Outside the limit the curvature parameter replaces the newtonian constant $G$ as the coupling parameter of Einstein's equation of gravitation. This new parameter may be approximately constant in certain physical conditions but, in general should be considered a variable which depends on the matter and energy on space-time. This effect and the geometric structure of matter and fields may be interpreted as "dark" matter effects. The substratum also determines a constant mass parameter for field excitations which obey the Dirac equation and behave as particles. Physically we consider that the substratum provides two related mass scales: the gravitational constant $G$ characterizes a macroscopic mass scale and a fundamental particle mass $m$ characterizes a microscopic mass scale. The substratum geometrically represents the physical concept of inertial system.

# 1. Introduction.

In a previous publication [1] it was indicated that a geometrical unified theory with SL(4,ℝ) as structure group produced the Einstein field equation of gravitation where the Einstein tensor $G_{\mu\nu}$ is determined by a geometric energy momentum tensor $\Theta_{\mu\nu}$. In addition the equation indicated that the gravitational constant was determined by the generalized nonriemannian curvature. In another paper [2] a constant odd nonriemannian connection solution was found. We discuss here how the nonriemannian curvature scalar plays the role of a gravitational coupling and leads the unification of different concepts of mass and to phenomena that may be interpreted as dark matter effects.

# 2. Fundamental Equations.

The first equation of the geometric theory is the field equation properly, a differential equation for the curvature $\Omega$ of the connection $\Gamma$, with a current source $J$ expressed in terms of the matter base $e$,

$$D^{*}\Omega = 4\pi\alpha \ ^{*}J = 4\pi\alpha e^{-1}\kappa e \ . \tag{1}$$

The second is the density of motion equation for the matter base, a generalized Dirac equation in terms of the covariant derivative $\nabla$ of the base,

$$\kappa^{\mu}\nabla_{\mu}e = \kappa^{\mu}\left(\partial_{\mu}e - e\Gamma_{\mu}\right) = -\tfrac{1}{2}\kappa^{\alpha}\nabla_{\mu}u^{\mu}_{\alpha}e = 0 \ . \tag{2}$$

The third is the energy momentum equation, which suggests the definition of a generalized energy momentum tensor and serves as an additional element to determine the orthonormal base and its torsion.

$$\mathrm{tr}\left[4\Omega_{\hat{\rho}\nu}\Omega^{\mu\nu} - u^{\mu}_{\hat{\rho}}\Omega^{\kappa\lambda}\Omega_{\kappa\lambda}\right] = k\,\mathrm{tr}\left[4e^{-1}\kappa^{\hat{\alpha}}u^{(\mu}_{\hat{\alpha}}\nabla_{\hat{\rho})}e - u^{\mu}_{\hat{\rho}}e^{-1}\kappa^{\hat{\alpha}}u^{\nu}_{\hat{\alpha}}\nabla_{\nu}e\right] \ . \tag{3}$$

Previous results [3] indicate that gravitation and electromagnetism are unified in a non trivial manner. There are additional generators which may represent the non classical interactions and produce effects related to nuclear energy and forces [4]. For spaces with matter, the energy momentum equation may be transformed into a generalized Einstein equation [1], which relates the Einstein tensor $G_{\mu\nu}$ to a total geometric energy momentum tensor $\Theta_{\mu\nu}$. Nevertheless there is an incorrect factor of 5/3 in [1], equation (2.19), due to the calculation of the geometric contributions to $\Theta$. In this paper we correct this numerical error.

For empty space the resultant equation, when the algebra reduces to its even part, is similar to the Stephenson [5] equation in the Yang's [6] theory. Fairchild [7] proved that this equation eliminates those solutions which are not, at the same time, Einstein's equation solutions with a cosmological constant. Therefore, in vacuum, we obtain the well known gravitational solutions.

For an internal solution with spherical symmetry the mass of a body may be defined in terms of energy or mass integrals, as in Einstein's theory. The corresponding external solution agrees with the standard tests of relativity.

# 3. The Comoving Substratum.

The proposed field equation (1) admits constant static solutions [2] for the connection and current fields. A rest state implies elimination of all kinetic effects. The remaining static effects, which do have importance, are determined by symmetries related to Killing vectors. We must measure excitations in a reference system comoving with the background matter or substratum, where the material base is always the identity. This is a generalization of the comoving coordinates systems used in general relativity. We assume that all time derivatives are zero. From the first fundamental differential equation we obtain an algebraic equation,

$$\omega^{c}_{\rho}\omega^{a\rho}\omega^{b\alpha}c_{abn}c_{cmd}g^{mn} = kJ^{\alpha}_{d} \tag{4}$$

We have a unique isotropic homogenous constant solution, proportional to the current, which we call the trivial solution. A Lorentz transformation determines an equivalent solution. We choose the name substratum for the equivalence class of these solutions under SO(3,1). In particular, we may write the substratum solution in the following form,

$$\omega = e^{-1}\left(-m_g dx^{\hat{\alpha}} \kappa_{\hat{\alpha}}\right)e + e^{-1}de = -\mathfrak{M}J + e^{-1}de = \Lambda + e^{-1}de \ . \tag{5}$$

The trivial connection is essentially proportional to the current, up to an automorphism. It is convenient to relate the proportionality constant $m_g$ to a fundamental geometric mass-energy $\mathfrak{M}$ constant defined by equation (5). It should be noted that in the expression for $\omega$, the term containing the current $J$ defines a potential tensorial form designated by $\Lambda$. Its subtraction from $\omega$ gives an object, $e^{-1}de$, that transforms as a connection. The tensorial form $\Lambda$ valued in the sl(4,$\mathbb{R}$) algebra may be expressed as

$$\Lambda = \frac{-\mathfrak{M}}{4}J \ . \tag{6}$$

The odd substratum connection determines an even nonriemannian curvature form,

$$\Omega = \omega \wedge \omega = m_g^2 J \wedge J = \frac{1}{16}\mathfrak{M}^2 dx^{\hat{\alpha}} \wedge dx^{\hat{\beta}} \kappa_{[\hat{\alpha}}\kappa_{\hat{\beta}]} \ . \tag{7}$$

Therefore, the curvature tensor of the substratum solution may be written as

$$\Omega^{\beta}{}_{\alpha\mu\nu} = \frac{1}{2}\mathfrak{M}^2 \delta^{\beta}_{[\mu}g_{\nu]\alpha} \tag{8}$$

showing the hyperbolic nature of the solution. The constant $\mathfrak{M}$ determines the curvature parameter of the substratum. The corresponding space to this curvature is conformally flat. The contracted curvature tensor is

$$\Omega_{\alpha\nu} = \frac{3}{4}\mathfrak{M}^2 g_{\alpha\nu} \ . \tag{9}$$

It is convenient to evaluate other energy terms related to the substratum solution. The energy momentum tensor is quadratic on the substratum curvature. Using eq. (8) we obtain

$$\Omega_{\alpha\beta\rho\nu}\Omega^{\alpha\beta}{}_{\mu}{}^{\nu} = 6g_{\rho\mu} \tag{10}$$

and the energy momentum tensor produced by the substratum curvature tensor is

$$^c\Theta = \Omega_{\alpha\beta\rho\nu}\Omega^{\alpha\beta}{}_{\mu}{}^{\nu} - \frac{1}{4}g_{\rho\mu}\Omega^{\alpha\beta\kappa\lambda}\Omega_{\alpha\beta\kappa\lambda} = 0 \ . \tag{11}$$

Similarly, the energy momentum tensor [1] produced by the substratum matter current is

$$^j\Theta_{\mu\rho} = -\frac{\alpha}{4}\mathrm{tr}\left[e^{-1}\kappa_{(\mu}\nabla_{\rho)}e - \tfrac{1}{4}g_{\mu\rho}e^{-1}\kappa^{\lambda}\nabla_{\lambda}e\right]$$
$$= -\frac{\alpha\mathfrak{M}}{16}\mathrm{tr}\left[\kappa_{(\mu}\kappa_{\rho)} + g_{\mu\rho}\right] = 0 \ . \tag{12}$$

Therefore, the substratum solution, which has an interaction energy $\Gamma.J$, establishes a zero level for the internal energies produced by the geometry and its matter source. This is consistent with the use of the substratum as a gravitational vacuum or as quantum vacuum.

The substratum solution is compatible with the energy equation and precisely provides a symmetric hyperbolic curvature necessary to obtain the newtonian limit. We shall consider that this limit is reached when the geometric equation solutions approach the substratum. In order to accomplish this, it is convenient to note that, for any connection solution $\Gamma$, we can always define a new connection by subtracting the tensorial potential form $\Lambda$ corresponding to the substratum solution,

$$\hat{\Gamma} \equiv \Gamma - \Lambda = \Gamma - \left(\frac{-\mathfrak{M}}{4}J\right) \ . \tag{13}$$

## 4. The Energy Momentum Equation.

The energy equation (3)

$$\text{tr}\left[4\Omega_{\hat{\rho}\nu}\Omega^{\mu\nu} - u_{\hat{\rho}}^{\mu}\Omega^{\kappa\lambda}\Omega_{\kappa\lambda}\right] = k\,\text{tr}\left[4e^{-1}\iota \circ u^{(\mu}\nabla_{\hat{\rho})}e - u_{\hat{\rho}}^{\mu}e^{-1}\iota \circ u^{\nu}\nabla_{\nu}e\right] \tag{14}$$

defines a tensor field on *M*. It is clear that the trace in the equation introduces a scalar product, the Cartan-Killing metric $^{C}g$, valued in the tensors. This Killing scalar product allows us to write the left-hand side of this equation in terms of a summation over all components along the 15 generators of a base in the sl(4,$\mathbb{R}$) Lie algebra.

It is necesary to split the tensors defined in [1], explicitly showing the components related to the different subalgebras and cosets. We shall denote by $^{L}\Omega$ the tensor associated to the sl(2,$\mathbb{C}$) subalgebra and by $^{c}\Omega$ the corresponding coset components. We decompose the curvature energy momentum tensor in the following manner.

$$\frac{1}{4}\text{tr}\left[\Omega_{\rho\nu}\Omega_{\mu}^{\ \nu} - \frac{1}{4}g_{\rho\mu}\Omega^{\kappa\lambda}\Omega_{\kappa\lambda}\right] = {^{c}g_{ab}}\left[{^{L}\Omega^{a}}_{\rho\nu}\,{^{L}\Omega^{b}}_{\mu}^{\ \nu} - \frac{1}{4}g_{\rho\mu}{^{L}\Omega^{a\kappa\lambda}}\,{^{L}\Omega^{b}}_{\kappa\lambda}\right] +$$
$$\phantom{\frac{1}{4}\text{tr}} {^{c}g_{\tilde{a}\tilde{b}}}\left[{^{c}\Omega^{\tilde{a}}}_{\rho\nu}\,{^{c}\Omega^{\tilde{b}}}_{\mu}^{\ \nu} - \frac{1}{4}g_{\rho\mu}{^{c}\Omega^{\tilde{a}\kappa\lambda}}\,{^{c}\Omega^{\tilde{b}}}_{\kappa\lambda}\right], \tag{15}$$

where the summation over the latin indices of $^{L}\Omega$ is restricted to the six components of the sl(2,$\mathbb{C}$) even subalgebra of sl(4,$\mathbb{R}$) and the summation over latin indices with tilde is over the nine components of the coset algebra. The latter terms, coming from the odd generators and the even $\kappa_5$ generator, correspond to the stress energy tensor component due to the additional nonriemannian "coset" fields present in the theory, including the standard electromagnetic field. They all have the familiar quadratic structure in terms of the curvature components,

$$^{c}\Theta_{\rho\mu} \equiv \frac{-1}{4\pi}\,{^{c}g_{\tilde{a}\tilde{b}}}\left[{^{c}\Omega^{\tilde{a}}}_{\rho\nu}\,{^{c}\Omega^{\tilde{b}}}_{\mu}^{\ \nu} - \frac{1}{4}g_{\rho\mu}{^{c}\Omega^{\tilde{a}\kappa\lambda}}\,{^{c}\Omega^{\tilde{b}}}_{\kappa\lambda}\right], \tag{16}$$

and define a coset field stress energy tensor $^{c}\Theta$. Since the electromagnetic generator is compact, the Killing metric introduces a minus sign and we must define $^{c}\Theta$ as shown so that the standard electromagnetic energy is positive definite.

The right hand side of the main equation (14) is interpreted as a stress energy tensor $^{j}\Theta$ related to the matter current source, in terms of the vector density $\iota$,

$$^{j}\Theta_{\hat{\rho}}^{\mu} \equiv \frac{-k}{4\pi}\,{^{c}g}\left(e^{-1}\iota \circ u^{(\mu}\nabla_{\hat{\rho})}e - \tfrac{1}{4}u_{\hat{\rho}}^{\mu}e^{-1}\iota \circ u^{\lambda}\nabla_{\lambda}e\right) =$$
$$-\frac{\alpha}{4}\text{tr}\left[e^{-1}\iota \circ u^{(\mu}\nabla_{\hat{\rho})}e - \tfrac{1}{4}u_{\hat{\rho}}^{\mu}e^{-1}\iota \circ u^{\lambda}\nabla_{\lambda}e\right]. \tag{17}$$

Thus, we may write equation (15) in this manner

$$^{c}g_{ab}\left[{^{L}\Omega^{a}}_{\rho\nu}\,{^{L}\Omega^{b}}_{\mu}^{\ \nu} - \frac{1}{4}g_{\rho\mu}{^{L}\Omega^{a\kappa\lambda}}\,{^{L}\Omega^{b}}_{\kappa\lambda}\right] - 4\pi\,{^{c}\Theta_{\rho\mu}} = -4\pi\,{^{j}\Theta_{\rho\mu}}. \tag{18}$$

The even sl(4,$\mathbb{R}$) generators corresponding to the left side in the last equation may be expressed in terms of the (2 to 1) homomorphic so(3,1) generators, 4×4 matrices $X_a$ (Lorentz rotation generators), acting on the tangent bundle *TM*. For these two algebras *L* and *L'* the respective Cartan-Killing metrics, defined by the trace over the dimension, differ by a factor of 2. Thus we may write

$$
\begin{aligned}
{}^C g_{ab} \left[ {}^L\Omega^a{}_{\rho\nu}\, {}^L\Omega^b{}_{\mu}{}^{\nu} - \frac{1}{4} g_{\rho\mu}\, {}^L\Omega^{a\kappa\lambda}\, {}^L\Omega^b{}_{\kappa\lambda} \right] &= 4\pi\, {}^L\Theta_{\rho\mu} \\
&= {}^C g'_{ab}(2) \left[ {}^{L'}\Omega^a{}_{\rho\nu}\, {}^{L'}\Omega^b{}_{\mu}{}^{\nu} - \frac{1}{4} g_{\rho\mu}\, {}^{L'}\Omega^{a\kappa\lambda}\, {}^{L'}\Omega^b{}_{\kappa\lambda} \right] \\
&= \frac{1}{4} \operatorname{tr}\left[ X_a X_b(2) \left( {}^{L'}\Omega^a{}_{\rho\nu}\, {}^{L'}\Omega^b{}_{\mu}{}^{\nu} - \frac{1}{4} g_{\rho\mu}\, {}^{L'}\Omega^{a\kappa\lambda}\, {}^{L'}\Omega^b{}_{\kappa\lambda} \right) \right] \\
&= \frac{1}{2} \operatorname{tr}\left( {}^{L'}\Omega^a{}_{\rho\nu}\, {}^{L'}\Omega^b{}_{\mu}{}^{\nu} - \frac{1}{4} g_{\rho\mu}\, {}^{L'}\Omega^{a\kappa\lambda}\, {}^{L'}\Omega^b{}_{\kappa\lambda} \right).
\end{aligned}
\quad (19)
$$

Furthermore, the curvature tensor has even sl(4,$\mathbb{R}$) components that arise from the quadratic product of components in the quotient sector of the connection and are clearly nonriemannian. The riemannian part of the curvature, designated by ${}^R\Omega$, is defined by the metric connection preserving the sl(4,$\mathbb{R}$) quadratic component and we may write in terms of a complementary nonriemannian part ${}^n\Omega$

$$
{}^{L'}\Omega^{\alpha}{}_{\beta\kappa\lambda} = {}^n\Omega^{\alpha}{}_{\beta\kappa\lambda} + {}^R\Omega^{\alpha}{}_{\beta\kappa\lambda}. \tag{20}
$$

Consequently, there is a contribution to the even stress energy tensor corresponding to this nonriemannian part of the curvature.

The connection associated to the SO(3,1) group, responsible for ${}^R\Omega$, admits the possibility of torsion. We can further split away the torsion from the Levi-Civita connection,

$$
\Gamma^{\alpha}_{\beta\mu} = \left\{ \begin{array}{c} \alpha \\ \beta\mu \end{array} \right\} + \Sigma^{\alpha}_{\beta\mu} \tag{21}
$$

and express the ${}^R\Omega$ curvature in terms of the Riemann tensor $R^{\alpha}{}_{\beta\mu\nu}$, defined by the space-time metric, and an explicit dependence on the torsion,

$$
{}^{L'}\Omega^{\alpha}{}_{\beta\kappa\lambda} = \left( {}^n\Omega^{\alpha}{}_{\beta\kappa\lambda} + R^{\alpha}{}_{\beta\kappa\lambda} \right) + Z^{\alpha}{}_{\beta\kappa\lambda} \equiv {}^nR^{\alpha}{}_{\beta\kappa\lambda} + Z^{\alpha}{}_{\beta\kappa\lambda}, \tag{22}
$$

where ${}^nR^{\alpha}{}_{\beta\mu\nu}$ is defined as a nonriemannian curvature including the Riemann tensor and

$$
Z^{\alpha}{}_{\beta\kappa\lambda} = \nabla_{\kappa}\Sigma^{\alpha}_{\beta\lambda} - \nabla_{\lambda}\Sigma^{\alpha}_{\beta\kappa} + \Sigma^{\alpha}_{\gamma\kappa}\Sigma^{\gamma}_{\beta\lambda} - \Sigma^{\alpha}_{\gamma\lambda}\Sigma^{\gamma}_{\beta\kappa}. \tag{23}
$$

It should be noticed that we define the Riemann tensor $R$ in the strict sense used originally by Riemann for metric spaces and denote by ${}^nR$ an even curvature that includes a nonriemannian part ${}^n\Omega$.

Substitution in equation (19) gives an expression in terms of the Riemann tensor of the symmetric metric connection,

$$
\begin{aligned}
4\pi\, {}^L\Theta_{\rho\mu} = & \frac{1}{2}\left( {}^nR^{\hat{\alpha}}{}_{\hat{\beta}\rho\nu}\, {}^nR^{\hat{\beta}}{}_{\hat{\alpha}\mu}{}^{\nu} - \frac{g_{\rho\mu}}{4}\, {}^nR^{\hat{\alpha}}{}_{\hat{\beta}}{}^{\kappa\lambda}\, {}^nR^{\hat{\beta}}{}_{\hat{\alpha}\kappa\lambda} \right) \\
& + \frac{1}{2} Z^{\hat{\alpha}}{}_{\hat{\beta}\rho\nu} Z^{\hat{\beta}}{}_{\hat{\alpha}\mu}{}^{\nu} - \frac{g_{\rho\mu}}{8} Z^{\hat{\alpha}}{}_{\hat{\beta}}{}^{\kappa\lambda} Z^{\hat{\alpha}}{}_{\hat{\beta}\kappa\lambda} + \frac{1}{2} Z^{\hat{\alpha}}{}_{\hat{\beta}\rho\nu}\, {}^nR^{\hat{\beta}}{}_{\hat{\alpha}\mu}{}^{\nu} - \frac{g_{\rho\mu}}{8} Z^{\hat{\alpha}}{}_{\hat{\beta}}{}^{\kappa\lambda}\, {}^nR^{\hat{\alpha}}{}_{\hat{\beta}\kappa\lambda} \\
& + \frac{1}{2}\, {}^nR^{\hat{\alpha}}{}_{\hat{\beta}\rho\nu} Z^{\hat{\beta}}{}_{\hat{\alpha}\mu}{}^{\nu} - \frac{g_{\rho\mu}}{8}\, {}^nR^{\hat{\alpha}}{}_{\hat{\beta}}{}^{\kappa\lambda} Z^{\hat{\alpha}}{}_{\hat{\beta}\kappa\lambda}.
\end{aligned}
\tag{24}
$$

The first term in parenthesis in the right hand side is similar to an expression previously developed by Stephenson [5] within the Yang [6] theory of gravitation,

$$H_{\rho\mu} = R^{\hat{\alpha}}{}_{\hat{\beta}\rho\nu} R^{\hat{\beta}}{}_{\hat{\alpha}\mu}{}^{\nu} - \frac{1}{4} g_{\rho\mu} R^{\hat{\alpha}}{}_{\hat{\beta}}{}^{\kappa\lambda} R^{\hat{\alpha}}{}_{\hat{\beta}\kappa\lambda} \quad . \tag{25}$$

We may define a stress energy tensor associated to the torsion

$${}^t\Theta_{\rho\mu} \equiv \frac{-1}{8\pi} \left( \begin{array}{c} Z^{\hat{\alpha}}{}_{\hat{\beta}\rho\nu} Z^{\hat{\beta}}{}_{\hat{\alpha}\mu}{}^{\nu} - \frac{g_{\rho\mu}}{4} Z^{\hat{\alpha}}{}_{\hat{\beta}}{}^{\kappa\lambda} Z^{\hat{\alpha}}{}_{\hat{\beta}\kappa\lambda} + Z^{\hat{\alpha}}{}_{\hat{\beta}\rho\nu} {}^n R^{\hat{\beta}}{}_{\hat{\alpha}\mu}{}^{\nu} - \frac{g_{\rho\mu}}{4} Z^{\hat{\alpha}}{}_{\hat{\beta}}{}^{\kappa\lambda} {}^n R^{\hat{\alpha}}{}_{\hat{\beta}\kappa\lambda} \\ + {}^n R^{\hat{\alpha}}{}_{\hat{\beta}\rho\nu} Z^{\hat{\beta}}{}_{\hat{\alpha}\mu}{}^{\nu} - \frac{g_{\rho\mu}}{4} {}^n R^{\hat{\alpha}}{}_{\hat{\beta}}{}^{\kappa\lambda} Z^{\hat{\alpha}}{}_{\hat{\beta}\kappa\lambda} \end{array} \right) \quad . \tag{26}$$

In this manner we may write equation (18) as

$$8\pi \left( -{}^c\Theta_{\rho\mu} - {}^t\Theta_{\rho\mu} \right) + {}^n H_{\rho\mu} = -8\pi \, {}^j\Theta_{\rho\mu} \tag{27}$$

and we *may* consider $H$ as the stress energy tensor of this geometric gravitation. The matter current energy momentum tensor ${}^j\Theta$ is equivalent to the *total* sum of geometric *field* energy momentum contributions. The *field* energy momentum tensor includes terms that are equivalent to the energy momentum associated to the motion of the *matter* current (particles). It is a question of convenience in the classical theories to consider this energy momentum to be *either* in the field *or* in the matter. In the geometric theory this equation is considered as an energy momentum balance equation rather than the proper field equation.

## 5. The Generalized Einstein Equation.

There is an alternate expression for $H$ obtained by decomposing the Riemann tensor in terms of the Weyl tensor, the Ricci tensor and the non riemannian curvature scalar ${}^n R$, using the expression

$${}^n R^{\beta\alpha}{}_{\mu\nu} = {}^n C^{\beta\alpha}{}_{\mu\nu} + 2\delta^{[\beta}_{[\mu} {}^n R^{\alpha]}_{\nu]} - \frac{1}{3} \delta^{[\beta}_{[\mu} \delta^{\alpha]}_{\nu]} {}^n R \quad . \tag{28}$$

Again we should note that we are dealing with the so(3,1) curvature tensor rather than the more general gl(4,r) curvature tensor also called Riemann by others. The quadratic term in $C$ vanishes [7],

$${}^n C^{\beta}{}_{\alpha\kappa\lambda} {}^n C^{\alpha}{}_{\beta\mu}{}^{\lambda} - \frac{1}{4} g_{\mu\kappa} {}^n C^{\beta}{}_{\alpha\nu\lambda} {}^n C^{\alpha}{}_{\beta}{}^{\nu\lambda} = 0 \quad , \tag{29}$$

and makes no contributions to $H$. The other contributions for the first term of ${}^n H$ become

$${}^n R^{\beta}{}_{\alpha\mu\nu} {}^n R^{\alpha}{}_{\beta\kappa}{}^{\nu} = -\frac{1}{3} {}^n R \, {}^n R_{\mu\kappa} - \frac{1}{2} g_{\mu\kappa} {}^n R^{\alpha\mu} {}^n R_{\alpha\mu} + \frac{1}{6} g_{\mu\kappa} {}^n R^2 - 2 \, {}^n R^{\alpha\lambda} {}^n C_{\alpha\mu\lambda\kappa} \tag{30}$$

obtaining finally in terms of the non riemannian curvature scalar ${}^n R$

$${}^n H_{\rho\mu} = \frac{-{}^n R}{3} \left( {}^n R_{\rho\mu} - \frac{1}{4} g_{\rho\mu} {}^n R \right) - 2 \, {}^n C^{\kappa}{}_{\mu\lambda\rho} {}^n R^{\lambda}_{\kappa} \quad . \tag{31}$$

The last expression implies that it may be formally written in terms of the Einstein tensor $G_{\mu\nu}$.

$${}^n H_{\rho\mu} = \frac{-{}^n R}{3} \left( G_{\rho\mu} + \frac{1}{4} g_{\rho\mu} R + {}^n \Omega_{\rho\mu} - \frac{1}{4} g_{\rho\mu} {}^n \Omega \right) - 2 \, {}^n C^{\kappa}{}_{\mu\lambda\rho} {}^n R^{\lambda}_{\kappa} \quad . \tag{32}$$

Because of the nonlinearity of the equations there is a contribution of gravitation to its own source. We may split $H$ into a $G$ dependent part, representing gravitation, and a complementary part, representing a geometric energy-momentum tensor contribution,

$$^g\Theta_{\mu\rho} \equiv \frac{1}{8\pi}\left[\frac{^nR}{3}\left(g_{\rho\mu}\frac{R}{4} + {^n\Omega_{\rho\mu}} - g_{\rho\mu}\frac{^n\Omega}{4}\right) + 2\,{^nC^\kappa}_{\mu\lambda\rho}\,{^nR^\lambda_\kappa}\right] \tag{33}$$

and represent equation (27) as Einstein's equation. The different $\Theta$ field terms may be grouped together; defining a total generalized geometric field energy momentum tensor for the external fields designated by $\Theta_{\mu\nu}$. Equation (14) may be written

$$\frac{^nR}{3}G_{\rho\mu} + 8\pi\left({^g\Theta_{\rho\mu}} + {^t\Theta_{\rho\mu}} + {^c\Theta_{\rho\mu}}\right) = 8\pi\,{^j\Theta_{\rho\mu}} \,. \tag{34}$$

We have a generalized Einstein equation with geometric stress energy tensors. Nevertheless, as in equation (14), the energy momentum tensor $^j\Theta$ of the matter current is equivalent to the *total* of geometric *field* energy momentum contributions including the Einstein tensor. If $^nR$ is nonzero we may write formally this Einstein equation,

$$G_{\rho\mu} = 8\pi\frac{3}{^nR}\left({^j\Theta_{\rho\mu}} - {^c\Theta_{\rho\mu}} - {^g\Theta_{\rho\mu}} - {^t\Theta_{\rho\mu}}\right) \equiv 8\pi\frac{3}{^nR}\Theta_{\rho\mu} \equiv 8\pi G T_{\rho\mu} \,, \tag{35}$$

which defines two stress energy tensors: the geometrical $\Theta$ and the classical $T$. The corresponding equation in [1] erroneously indicates a factor of 5 instead of the factor of 3 in the numerator in the equation. The geometric energy momentum tensor has terms which reflect the energy and motion of matter and interaction potentials in a similar way to known physical situations. Nevertheless, it is possible that it includes unknown geometric terms which may be related to the so called dark matter and energy. In certain phenomenological macroscopic situations it is also possible that this tensor approaches only a combination of the tensors normally used in astrophysics. In any case, the fundamental gravitational difference rests in the presence of the nonriemannian curvature scalar $^nR$ in place of $G$ in the equation, caused by its quadratic structure.

As said before, the relation of the geometric stress-energy tensor density field $\Theta$ with the standard physical density $\rho$ obtained in the newtonian limit provides a physical determination of an indeterminate parameter relating macroscopic and microscopic mass scales. In other words, it is possible to equate $\mathfrak{M}^{-2}$ to the value of the gravitational constant and $m$ to the value of a fundamental particle mass without contradictions. Physically we consider that the substratum provides two conceptually related mass scales: in the fundamental representation it determines the gravitational parameter $G$ that characterizes a macroscopic mass scale and in the induced representation it determines a fundamental particle mass $m$ that characterizes a microscopic mass scale.

It should be kept in mind that, as we said before, that this equation should be considered an energy momentum balance equation rather than the proper field equation. Nevertheless, the conservation of the tensor $G$ with respect to the induced Levi-Civita connection in the bundle $TM$ implies the conservation of a tensor defined by

$$\nabla_\rho G^{\rho\mu} = 8\pi\nabla_\rho\left(\frac{3\Theta^{\rho\mu}}{^nR}\right) = 0 \,. \tag{36}$$

There should be compatibility of the resultant equations with those obtained from the conservation of the current $J$, which determine equations of motion, [8, 9]. If the stress energy tensor is decomposed in terms of a multipole expansion we find the usual equations of motion [11]: geodesic equation for a monopole, equations for a spinning particle and other multipole equations of motion.

Using eq. (13) the even so(3.1) curvature component may be split into the substratum part $^s\Omega$, determined by eq. (9) in terms of the new metric, the Riemann tensor of the new connection and a complementary nonriemannian part $^n\Omega$ determined only by the total odd connection,

$$^nR^\alpha{}_{\beta\kappa\lambda} = {^n\widehat{\Omega}^\alpha{}_{\beta\kappa\lambda}} + \widehat{R}^\alpha{}_{\beta\kappa\lambda} + {^s\Omega^\alpha{}_{\beta\kappa\lambda}} \,. \tag{37}$$

Einstein's equation (35) may be written as

$$G_{\rho\mu} = 8\pi \frac{3}{^nR}\Theta_{\rho\mu} = 8\pi \frac{3}{3\mathfrak{M}^2 + {^n\widehat{\Omega}} + \widehat{R}}\Theta_{\rho\mu} \equiv 8\pi G T_{\rho\mu} \; . \tag{38}$$

The only nonriemannian curvature contribution due to the substratum is through the scalar $^nR$. All other contributions of the conformally flat curvature cancel themselves. The expression of the scalar field $^nR$ obtained using eq. (37) in terms of corresponding curvature scalar fields shows that the gravitational equation has a variable factor that may be included as a modified effective energy momentum tensor $T$. This matter tensor may be written in terms of the constant $\mathfrak{M}$.

The introduction of the gravitational parameter $G$ in the scalar field $^nR$ modifies the Einstein equation by a factor that may be interpreted as an "apparent" increment (or decrement) of an effective energy momentum tensor. The Einstein tensor may be written

$$G_{\rho\mu} = \frac{8\pi\Theta_{\rho\mu}}{\mathfrak{M}^2 + \left({^n\widehat{\Omega}} + \widehat{R}\right)/3} = \frac{8\pi}{\mathfrak{M}^2}\frac{\Theta_{\rho\mu}}{1 + \left({^n\widehat{\Omega}} + \widehat{R}\right)/3\mathfrak{M}^2} = 8\pi G \frac{\Theta_{\rho\mu}}{1 + G\left({^n\widehat{\Omega}} + \widehat{R}\right)/3} \; . \tag{39}$$

In general solutions, as shown in equation (35), the even non riemannian curvature scalar $^nR$ plays the role of gravitational coupling. In solutions near the substratum solution the scalar $^nR/3$ should have a value near the constant $G$, which would be the exact value in the newtonian limit. This increment may be responsible for the detection of dark matter and energy. In particular, the Friedmann cosmological metric

$$d\tau^2 = dt^2 - a^2(t)\left(d\chi^2 + \Sigma\left(d\theta^2 + \sin^2\theta d\phi^2\right)\right) \quad \begin{cases} k=1 & \Sigma = \sin\chi \\ k=0 & \Sigma = \chi \\ k=-1 & \Sigma = \sinh\chi \end{cases} \tag{40}$$

would correspond to an equation for the Einstein tensor which may be written as follows, using a well known result [11] for its scalar curvature $R$ and neglecting $\Omega$,

$$G_{\rho\mu} = 8\pi G \frac{\Theta_{\rho\mu}}{1 - 2G\left(\frac{\dot{a}^2 + k}{a^2} + \frac{\ddot{a}}{a}\right)} \; . \tag{41}$$

Of course, the new solutions for the radius $a$ of the universe, using the different known energy momentum tensors, will differ from the presently known solutions. This geometric equation may be interpreted as the existence of an apparent effective stress energy tensor which includes contributions from "dark matter",

$$T_{\rho\mu} = \frac{\Theta_{\rho\mu}}{1 + GR/3} = \frac{\Theta_{\rho\mu}}{1 - 2G\left(\frac{\dot{a}^2 + k}{a^2} + \frac{\ddot{a}}{a}\right)} \; . \tag{42}$$

For the case of a pure metric gravitation theory there are zero coset fields $^c\Theta$ and zero torsion $\Sigma$. We have then

$$G_{\rho\mu} = 8\pi\frac{3}{R}\left({^j\Theta_{\rho\mu}} - {^g\Theta_{\rho\mu}}\right) \equiv 8\pi\frac{3}{R}{^m\Theta_{\rho\mu}} \tag{43}$$

where we have defined the matter current energy momentum tensor $^m\Theta$.

In vacuum, additionally there is no matter current. Only $^g\Theta$ remains and we get back the Stephenson-Yang equation for vacuum,

$$H_{\rho\mu} = \frac{-R}{3}\left(R_{\rho\mu} - \frac{1}{4}g_{\rho\mu}R\right) - 2C^{\kappa}{}_{\mu\lambda\rho}R^{\lambda}_{\kappa} = 0 \quad . \tag{44}$$

Fairchild has shown [7] that a null $H_{\mu\nu}$ implies, in Yang's theory, that the only empty space solutions are the Einstein spaces, ruling out the exceptional static spherically symmetric solutions given by Thomson [12] and Pavelle [13, 14]. A simple way to prove this here is to decompose $R_{\mu\nu}$ into its trace and traceless parts.

$$R_{\rho\mu} = \frac{R}{4}g_{\rho\mu} + P_{\rho\mu} \tag{45}$$

and substitute into (43)

$$H_{\rho\mu} = \frac{-R}{3}P_{\rho\mu} - 2C^{\kappa\,\lambda}{}_{\mu\,\rho}P_{\lambda\kappa} = 0 \quad . \tag{46}$$

This equation may be written as an eigenvalue equation for an operator *C* with eigenvector *P*

$$CP = \left(\frac{-R}{6}\right)P \quad . \tag{47}$$

The traceless symmetric *P* is spanned by a *9* dimensional linear space. The set of diagonal components of the 9×9 *C* matrix operator vanishes in any real coordinate system because of the properties of the Weyl tensor. On the other hand, in order to reproduce a nonzero eigenvector by the action of the operator, it is necessary that there exists a real coordinate system where *C* is a nonzero diagonal matrix. Since no such system exists the only solution of eq. (47) is that the eigenvector *P* is the zero vector. The expression in equation (45), with zero *P* defines the Einstein spaces. These spaces clearly satisfy equation (43) and therefore are the only possible pure gravitational vacuum solutions in this theory. They correspond to the Einstein equation with a cosmological constant.

## 6. The Newtonian Limit and the *G* Constant.

It is usual to assume, as a newtonian approximation, that the characteristic parameters of a newtonian solution of Einstein's equation $v/c \ll 1$, $\varphi \ll 1$ are of order $\varepsilon^2$ [15] in a small dimensionless parameter $\varepsilon$. To obtain the newtonian limit of the geometric theory we use a geometric definition of a space-time limit first given by Geroch [16] for metric spaces and generalized to spaces with a connection [17]. The geometric formulation of Newton's gravitation is the Newton-Cartan theory [18, 19, 20]. The main mathematical difficulty in obtaining the limit is that this formulation does not provide a metric but relies on a nonriemannian affine connection, a tensor of valence (2,0) and rank 3 (singular metric) and a scalar time function.

This small parameter $\varepsilon$ may be related to the orthonormal tetrad *u*, thus characterizing the propagation of gravitational disturbances. The newtonian limit of Einstein type theories of gravitation is discussed in detail in [17]. There it is shown that, in the limit $\varepsilon \to 0$, the metric becomes singular in $\varepsilon$. Nevertheless the connection remains regular in the limit and defines a newtonian affine connection not related to a metric. Since we have taken the connection as the fundamental representation of gravitation, the gravitational limit may be defined appropriately. The corresponding newtonian curvature tensor is the limit of the Riemann tensor. The projection of this tensor on the tridimensional time hypersurfaces *t* defines a Riemann tensor on them, which is not necessarily flat.

Nevertheless, assumptions on the stress energy tensor determine that the tridimensional Newton space is flat. In this case the nonvanishing components ${}^0\Gamma_{00}{}^a$ of the limit connection would give the only nonvanishing components of the curvature tensor determining Poisson's equation.

In the geometric theory, the newtonian limit $\varepsilon \to 0$ should be obtained from eq. (35). The curvature scalar *R* may become singular in the limit due to the singularity in the metric. It is possible to make assumptions on the geometry to avoid this singularity, but it is also possible to let the geometry be determined by the stress energy tensor, making assumptions of regularity for the latter tensor. In Einstein's theory this singularity may be handled by actually moving it to the stress energy tensor

$$-R = g^{\mu\nu}\left(R_{\mu\nu} - \frac{1}{2}g_{\mu\nu}R\right) = \kappa g^{\mu\nu}T_{\mu\nu} = \kappa T \quad, \tag{48}$$

$$R_{\mu\nu} = \kappa\left(T_{\mu\nu} - \frac{1}{2}g_{\mu\nu}T\right) \quad. \tag{49}$$

In the limit the relation between the scalars $R$ and $T$ is broken due to singularity in the metric. When the newtonian limit is taken, it is assumed that the stress energy tensor has such properties that the possible singularity in $R$ is avoided.

We actually may do the same to handle the possible $R$ singularity present in the Einstein tensor $G$ adjoining it to $\Theta$. Equivalently it is also possible to move the term from the left side to the right side. We also assume, as in Einstein's theory, that the contravariant stress energy tensor remains regular in the limit. As shown [17] the curvature tensor in the limit, in terms of a hypersurface orthogonal timelike vector, is

$$^{0}R_{\mu\nu} = \lim_{\varepsilon \to 0}\left(\frac{1}{2}\kappa T^{\alpha\beta}t_{\alpha}t_{\beta}t_{\mu}t_{\nu} + \mathcal{O}(\varepsilon^2)\right) = \frac{1}{2}\kappa\,{}^{0}T^{\alpha\beta}t_{\alpha}t_{\beta}t_{\mu}t_{\nu} \quad. \tag{50}$$

As in Einstein's theory, the curved time hypersurfaces become flat in the limit because the field equation

$$^{0}\overline{R}_{mn} = \lim_{\varepsilon \to 0}\overline{R}_{mn} = \lim_{\varepsilon \to 0}\kappa\left(g_{m\alpha}g_{n\beta}T^{\alpha\beta} - \frac{1}{2}g_{mn}g_{\alpha\beta}T^{\alpha\beta}\right) = \lim_{\varepsilon \to 0}\mathcal{O}(\varepsilon^2) = 0 \tag{51}$$

for the tridimensional curvature determines that the corresponding tridimensional Riemann tensor $^{0}\overline{R}$ is zero in the limit. Similarly the other space components of the Riemann tensor $R$ also vanish and the time vector $t^{\mu}$ becomes orthogonal to these hypersurfaces.

When establishing this limit, for any connection solution $\Gamma$, we can always define a new connection by subtracting the tensorial potential form $\Lambda$ corresponding to the substratum solution as indicated in eq. (13),

$$\widehat{\Gamma} \equiv \Gamma - \Lambda = \Gamma - \left(\frac{-\mathcal{M}}{4}J\right) \quad. \tag{52}$$

The even so(3.1) curvature component may be split into the substratum part $^{s}\Omega$, determined by eq. (9) in terms of the new metric, the Riemann tensor of the new connection and a complementary nonriemannian part $^{n}\Omega$ determined only by the total odd connection,

$$^{n}R^{\alpha}{}_{\beta\kappa\lambda} = {}^{n}\widehat{\Omega}^{\alpha}{}_{\beta\kappa\lambda} + \widehat{R}^{\alpha}{}_{\beta\kappa\lambda} + {}^{s}\Omega^{\alpha}{}_{\beta\kappa\lambda} \quad. \tag{53}$$

Einstein's equation (35) may be written as

$$G_{\rho\mu} = 8\pi\frac{3}{{}^{n}R}\Theta_{\rho\mu} = 8\pi\frac{3}{3\mathcal{M}^2 + {}^{n}\widehat{\Omega} + \widehat{R}}\Theta_{\rho\mu} \equiv 8\pi G T_{\rho\mu} \tag{54}$$

The only nonriemannian curvature contribution due to the substratum is through the scalar $^{n}R$. All other contributions of the conformally flat curvature cancel themselves.

Since Newton's theory is a pure gravitational theory, we assume that the new connection, as we approach the limit, is strictly gravitational. It is necessary to make the previous assumptions of regularity of the matter stress energy tensor. When we take the newtonian limit of the generalized Einstein equation (35), the three dimensional space curvature vanishes as shown in eq. (51). Then the only surviving component of the Riemann tensor satisfies the equation

$$\widehat{R}_{00} = \frac{8\pi}{2}\lim_{\varepsilon \to 0}\left(\frac{3\Theta_{\hat{0}\hat{0}}}{{}^{n}R}\right) = 4\pi\lim_{\varepsilon \to 0}\left(\frac{3\Theta_{\hat{0}\hat{0}}}{{}^{s}\Omega + {}^{n}\widehat{\Omega} + \widehat{R}}\right) \quad. \tag{55}$$

If in the newtonian limit we simultaneously approach the trivial substratum solution, $R$ and ${}^n\Omega$ are negligible, in the limit, with respect to the constant ${}^s\Omega$. The limit of ${}^nR$ is

$$\lim_{\varepsilon \to 0} {}^nR = \lim_{\varepsilon \to 0}\left({}^nR^\alpha{}_{\beta\kappa\alpha}g^{\beta\kappa}\right) = \lim_{\varepsilon \to 0}\left[\left({}^n\widehat{\Omega}^\alpha{}_{\beta\kappa\alpha} + \widehat{R}^\alpha{}_{\beta\kappa\alpha} + {}^s\Omega^\alpha{}_{\beta\kappa\alpha}\right)g^{\beta\kappa}\right]$$
$$= \lim_{\varepsilon \to 0}\left[\frac{3\mathcal{M}^2}{4}g_{\beta\kappa}g^{\beta\kappa}\right] = \frac{3\mathcal{M}^2}{4}\lim_{\varepsilon \to 0}\left[1 + \varepsilon^2 h_{bk}\left(\frac{h^{bk}}{\varepsilon^2}\right)\right] = 3\mathcal{M}^2 \quad . \tag{56}$$

The field equation in the newtonian limit may then be written in the following form:

$$\partial_a\partial^a\varphi = 4\pi\lim_{\varepsilon \to 0}\left(\frac{\Theta_{\hat{0}\hat{0}}}{\mathcal{M}^2}\right) = \frac{4\pi}{\mathcal{M}^2}\lim_{\varepsilon \to 0}\Theta_{00} \equiv 4\pi G\rho \tag{57}$$

and the field equation becomes the Poisson equation and defines the value of Newton's constant $G$ as a substratum parameter.

In the newtonian limit the geometric and classic densities are related by the constants $\mathcal{M}$ and $G$. In general solutions, as shown in equation (35), the even curvature scalar ${}^nR$ plays the role of a coupling function. In solutions near the substratum solution the scalar ${}^nR/3$ should have a value near the constant $G$, which is the exact value of the coupling constant in the newtonian limit. If the odd connection nonriemannian fields contribute to the curvature scalar ${}^nR$, the coupling parameter would be variable, diminishing with the field strength. This effect may be interpreted as the presence of dark matter.

For example, assume a body of mass $M$ produces a Friedmann metric as indicated in section 5 and choose solution parameters so that $k$ is zero and there is no acceleration. Using eqs. (41) or (42) the newtonian velocity of a body rotating under this field is predicted by the physical geometry to be

$$v^2 = \frac{MG}{r\left(1 - 2G\left(\dot{a}/a\right)^2\right)} = \frac{v_0^2}{1 - 2G\left(\dot{a}/a\right)^2} \quad , \tag{58}$$

where $v_0$ is the velocity of the particle if the Friedmann metric velocity were to be zero. This pure geometric result that increases the apparent rotational velocity produces astrophysical effects which may also be blamed on "dark matter".

# 7. Physical Significance of the Comoving Substratum.

We have seen that for a pure gravitational spherical solution, the Schwarzschild mass $\mathcal{M}$ is the integral of the energy density using this ${}^nR/3$ coupling, as shown in [1]. This mass $\mathcal{M}$ determines the Schwarzschild geometry geodesics and the particle motion without explicit knowledge of a parameter $G$. In principle, similar relations should exist for other solutions. The matter density expressed in equation (35) is in agreement with the definition of this total mass $\mathcal{M}$. In other words the general density of matter, corresponds to the source of Poisson's equation in Newton´s theory.

There is a fundamental scalar interaction energy field on the substratum characterized by the constant mass-energy parameter $\mathfrak{M}$, corresponding to the mass definition from $J$ and $\omega$, using the Cartan-Killing metric ${}^cg$ in the defining representation. This scalar energy field is the dual of a density 4-form $\omega \wedge {}^*J$ field also characterized by $\mathfrak{m}$. There also is a current stress-energy tensor density field ${}^j\Theta$ on the substratum related to $J$ and $\omega$. Essentially the geometric objects on the substratum are determined by these two $A$-valued forms: the current 3-form $*J$ and the connection 1-form $\omega$. These forms are constant on the substratum.

Since the Cartan-Killing metric depends on the particular representation [21], the fundamental mass-energy parameter that characterizes the substratum is determined up to a representation dependent normalization constant. For a representation $\mathcal{D}(A)$ the mass parameter is, with the normalization we are using,

$$m = \frac{\operatorname{tr} J^\mu \Gamma_\mu}{\operatorname{tr} I} = \frac{\operatorname{tr} J^\mu \Gamma_\mu}{\operatorname{tr}\left(-\kappa^0 \kappa_0\right)} = \frac{\operatorname{tr}\left(\mathcal{D}(J)\bullet\mathcal{D}(\Gamma)\right)}{\operatorname{tr}\left(-\mathcal{D}(\kappa^0)\mathcal{D}(\kappa_0)\right)}$$
$$= \frac{\operatorname{tr}\left(\mathcal{D}(J)\bullet\mathcal{D}(\Gamma)\right)}{N_\mathcal{D} \operatorname{tr} \mathcal{D}(I)} = \frac{\operatorname{tr}\left(\mathcal{D}(J)\bullet\mathcal{D}(\Gamma)\right)}{N_\mathcal{D} d_\mathcal{D}} \tag{59}$$

in terms of the dimension $d_\mathcal{D}$ and a constant $N_\mathcal{D}$, both related to the representation $\mathcal{D}$.

Apart from the defining representation, we have to consider the induced representations [22] necessary to characterize particle excitations. This question has been discussed in detail in other publications [23]. The value of the mass parameter in these representations corresponds to eigenvalues of the fundamental equation (2) after substitution of the connection using equation (13) and is related to the value $\mathcal{M}$ obtained in the defining representation. Equation (2) becomes

$$\kappa^\mu \nabla_\mu e = \kappa^\mu \left(\partial_\mu e - e\, {}^0\Gamma_\mu\right) = \kappa^\mu \left[\partial_\mu e - e\hat{\Gamma}_\mu - e\left(\frac{-m}{4} J_\mu\right)\right] = \kappa^\mu \hat{\nabla}_\mu e - m e = 0 \tag{60}$$

which essentially is the Dirac equation with the connection acting as potential and shows that the bare mass $m$ associated to the excitation (particle), in the corresponding representation, is also determined by the constant mass-energy parameter $\mathcal{M}$.

This $m$ is a fundamental Dirac physical mass parameter, associated to a particular induced representation, and should be related by a constant to the interaction energy scalar parameter $\mathcal{M}$,

$$m = \mathcal{M}/N \ . \tag{61}$$

In order to find this constant $N$ it is sufficient to calculate the Cartan-Killing product for one generator in both representations [21]. The constant, that relates the parameters in the defining matrix representation and the induced representation, is mathematically an indeterminate number because it depends on the ratio of infinite value integrals on the symmetric noncompat coset spaces involved in the Cartan-Killing metric in the induced representation. The relation of the geometric stress-energy tensor density field $\Theta$ with the standard physical density $\rho$ expressed by equation (57) obtained in the newtonian limit, provides a physical determination of this indeterminate constant. In other words, it is possible to equate $\mathcal{M}^{-2}$ to the value of the gravitational constant and $m$ to the value of a fundamental particle mass without contradictions. Physically we consider that the substratum provides two conceptually related mass scales: in the fundamental representation it determines the gravitational parameter $G$ that characterizes a macroscopic mass scale and in the induced representation it determines a fundamental particle mass $m$ that characterizes a microscopic mass scale. In fact the different notions of inertial mass, gravitational mass, active mass and passive mass are unified in terms of substratum energy.

The riemannian curvature associated to the even part of $\Omega$ may correspond to a symmetric tridimensional space that was used by Einstein in cosmological applications. From this cosmological gravitational point of view, the space-like tridimensional manifold is a symmetric spaces $S^3$, $H^3$ or $R^3$. Assuming the $S^3$ symmetry, Einstein [24] derived a relation between the total mass $M$, the radius of the universe $\mathcal{R}$ and the gravitational parameter. Putting the results together we obtain an estimate for the substratum mass-energy parameter $\mathcal{M}$ which represents the radial line mass density,

$$\mathcal{M} = \sqrt{2M/\pi\mathcal{R}} \ . \tag{62}$$

Therefore, the substratum interaction energy density is related to mass in different contexts. In particular the energy determines the following: 1- the Dirac particle mass parameter $m$, in corresponding representations; 2- the Schwarzschild metric gravitational mass parameter $\mathcal{M}$; 3- the gravitational parameter $G$. The concept of macroscopic active gravitational mass is consistent with the concept of microscopic inertial mass.

## 8. Conclusions.

The substratum nonriemannian curvature scalar is determined by the energy-mass content of the interaction of the connection field and the mater frame. This parameter is related to Newton's gravitational constant in the newtonian limit. Nevertheless in general solutions the curvature scalar is a parameter of the generalized Einstein equation of gravitation. This scalar may be approximately constant in certain physical conditions when Newton's theory applies, but in general should be considered a variable which depends on the matter and energy on space-time. This effect and the geometric structure of matter may be interpreted as "dark" matter.

The even curvature scalar $\Omega$ may be considered the germ and origin of the physical constants related to the substratum and, in general, plays this role in nonconstant solutions. Physically we consider that the substratum provides two related mass scales: the gravitational constant $G$ characterizes a macroscopic mass scale and a fundamental particle mass $m$ characterizes a microscopic mass scale, in accordance with Mach's ideas.

The substratum is the background around which the particle excitations of frame and connection fields are. It provides a mechanism for assigning constant masses to these particles. The equation of motion may be written as a Dirac equation in terms of these constant masses. In general, the substratum determines the inertial properties of matter. In this sense we may say that it geometrically represents the physical concept of inertial system.

The general relativistic Einstein equations with cosmological constant follow from the stress energy equation for empty space. For nonempty space we obtain a generalized Einstein equation (35) relating the Einstein tensor $G_{\mu\nu}$ to a geometric stress energy tensor $\Theta_{\mu\nu}$. For an internal solution with spherical symmetry the mass of a body may be defined in terms of energy-mass integrals, as in Einstein's theory. The matching exterior solution is in agreement with the standard relativity tests. Further, there exists the newtonian limit where we obtain the corresponding Poisson equation in terms of a geometric energy density related to the gravitational constant $G$ or the characteristic Planck length $\lambda$

On the substratum there is a fundamental scalar interaction energy field characterized by the constant mass-energy parameter $\mathfrak{M}$, corresponding to the mass definition from $J$ and $\omega$, using the Cartan-Killing metric $^c g$ in the defining representation. This scalar energy field is the dual of a density 4-form field also characterized by $\mathfrak{M}$. The current stress-energy tensor density field $^j\Theta$ on the substratum is also related to $J$ and $\omega$. Essentially the geometric objects on the substratum are determined by these two $A$-valued forms: the current 3-form *$J$ and the connection 1-form $\omega$. These forms are constant on the substratum.